\documentstyle[preprint,aps]{revtex}
\newcommand{\be}{\begin{eqnarray}}
\newcommand{\ee}{\end{eqnarray}}

\begin{document}

\draft
\preprint{LBL-37105}
%\preprint{April 1995}
\title{Phenomenology of Charm and Bottom Production$^\star$}
\footnotetext{$^\star$ This work was supported in part by the Director,
Office of Energy Research, Division of Nuclear Physics of the Office of High
Energy and Nuclear Physics of the U. S. Department of Energy under Contract
Number DE-AC03-76SF0098.}

\author{R. Vogt$^\dagger$}
\footnotetext{$^\dagger$ For detailed numerical results, contact
vogt@nsdssd.lbl.gov.}
\address{Nuclear Science Division, Lawrence Berkeley Laboratory,
Berkeley, CA 94720, USA\\
and\\
Physics Department, University of California at Davis, Davis, CA 95616 USA}

\maketitle

\begin{abstract}
We discuss the renormalization and factorization scale
dependence of charm and bottom production both at fixed-target energies and
at present and future colliders.  We investigate whether distributions
calculable at leading order can be extrapolated to next-to-leading
order by a constant multiplicative factor.
\end{abstract}
\newpage

Heavy quark hadroproduction is currently being studied both at fixed-target
energies with proton and pion beams and at collider energies.  Additionally,
charm and bottom quark production will be copious at the future RHIC and LHC
colliders where nucleus-nucleus
collisions are expected to produce a deconfined state
of quarks and gluons, the quark-gluon plasma.  Signatures of deconfinement
include lepton pair production by partons in the plasma \cite{VJMR} and an
enhancement of heavy quark production \cite{Shor,Shuryak}.  Thus a
detailed knowledge of heavy quark production is required to separate its
production during the initial nucleon-nucleon collisions from that by the
quark-gluon plasma.

Previous measurements of the $c \overline c$ production cross section
at $\sqrt{S} \leq 63$ GeV suggested that
the lowest order (Born) cross section underpredicted the data by a factor of
two to three \cite{Reu,Appel},
called the $K$ factor after a similar situation in Drell-Yan
production.  More generally,
\begin{equation} K_{\rm exp} = \frac{\sigma_{\rm data}(AB
\rightarrow Q \overline Q)}{\sigma_{\rm theory}(AB \rightarrow Q \overline Q)}
\, \, , \end{equation} where $Q$ is the produced heavy quark.
The projectile and target, $A$ and $B$, can be either
hadrons or nuclei.  The next-to-leading order (NLO)
corrections to the Born cross section have
been calculated \cite{NDE1,NDE2,SvN} and an analogous
theoretical $K$ factor can be
defined from the ratio of the NLO to the Born cross sections, \begin{equation}
K_{\rm th} = \frac{\sigma_{\rm NLO}(AB
\rightarrow Q \overline Q)}{\sigma_{\rm Born}(AB \rightarrow Q \overline Q)}
\, \, , \end{equation} where $\sigma_{\rm NLO}$ is the sum of the Born and
${\cal O}(\alpha_s)$ corrections.  Particularly
for the ``lighter'' heavy quarks, $c$ and $b$,
the NLO cross section is strongly dependent
on the choice of the renormalization and factorization scales
which determine both $K_{\rm exp}$ and $K_{\rm th}$.

In this paper we
discuss the scale dependence of $c$ and $b$ quark production and its
influence on the $K$ factors. While we explore the dependence
within the limits of perturbative QCD, we adjust the scales and
the heavy quark mass, $m_Q$, to achieve $K_{\rm exp}^{\rm NLO} \approx
1$\footnote{$K_{\rm exp}^{\rm NLO}$ is the $K$ factor for the data compared
to the NLO theory calculation.}
\cite{HPC} keeping in mind that further corrections to
the cross section could also be large. We use these mass and scale
parameters to predict the cross sections and $K_{\rm th}$
for charm and bottom production at RHIC and LHC energies, $\sqrt{S} = 200$-500
GeV and 5.5-14 TeV respectively.
Since only the Born contribution is often used in event generators, it is
important to check that
the distributions calculable at the Born level are a good
approximation to the NLO
results.  We show that $K_{\rm th}$ is nearly constant
if the renormalization and factorization scales are
assumed to be a function of the heavy quark transverse momentum, $p_T$.
Our calculations are done with a Monte Carlo
program developed by Nason and collaborators \cite{NDE1,NDE2,MNR}.

The double differential cross section
for $Q \overline Q$ pair production by hadrons $A$ and $B$
is \be E_Q E_{\overline Q} \frac{d\sigma_{AB}}{d^3p_Q d^3p_{\overline Q}}
= \sum_{i,j}\int \,dx_1\, dx_2 F_i^A(x_1,\mu_F)F_j^B(x_2,\mu_F)
E_Q E_{\overline Q} \frac{d\widehat{\sigma}_{ij}
(x_1P_1,x_2P_2,m_Q,\mu_F,\mu_R)}{d^3p_Q d^3p_{\overline Q}} \, \,  . \ee
Here $i$ and $j$ are the interacting
partons and the functions $F_i$ are the number densities of gluons, light
quarks and antiquarks ($m<m_Q$)
evaluated at momentum fraction $x$ and factorization
scale $\mu_F$. The short-distance cross section,
$\widehat{\sigma}_{ij}$, is calculable as a perturbation series in
$\alpha_s(\mu_R)$ where the strong coupling constant is evaluated at
the renormalization scale $\mu_R$.  Both $\mu_F$ and $\mu_R$
are of the order of the heavy quark mass.

At the Born level, ${\cal O}(\alpha_s^2)$,
the cross section can be written as
\be E_Q E_{\overline Q} \frac{d\sigma_{AB}}{d^3p_Q d^3p_{\overline Q}}
= \int \frac{s}{2 \pi} \,dx_1\, dx_2 \, C(x_1,x_2) \, \delta^4(x_1P_1
+ x_2P_2 - p_Q - p_{\overline Q}) \, \, \ee
where $\sqrt{s}$, the
parton-parton center-of-mass energy, is related to $\sqrt{S}$, the
hadron-hadron center-of-mass energy, by $s = x_1 x_2 S \geq 4m_Q^2$.
The intrinsic transverse momenta of the incoming partons have been neglected.
The convolution of the subprocess
cross sections with the parton number densities is contained in
$C(x_1,x_2)$, \be C(x_1,x_2) = \sum_q
[F_q^A(x_1)  F_{\overline q}^B(x_2) +  F_{\overline q}^A(x_1) F_q^B(x_2)]
\frac{d
\widehat{\sigma}_{q \overline q}}{dt} + F_g^A(x_1) F_g^B(x_2)
\frac{d \widehat{\sigma}_{gg}}{dt} \, \, , \ee
where production is by $q \overline q$ annihilation,
$ q \overline q \rightarrow  Q \overline Q$, and gluon fusion,
$g g \rightarrow Q \overline Q$.
The scale dependence has been suppressed since there is no distinction between
$\mu_F$ and $\mu_R$ at this order.  The scale
$2m_Q$ is commonly used for the total cross section, motivated by
the $s$-channel processes.  However, the
$t$ and $u$ channel gluon fusion graphs
involve heavy quark exchange between the gluons, suggesting
that $m_Q$ is a better scale choice.  Thus some ambiguity in scale
already exists at leading order.

Four-momentum conservation leads to the rather simple expression
\be \frac{d \sigma_{AB}}{dp_T^2 dy_Q dy_{\overline Q}} = x_1 x_2
C(x_1,x_2) \, \, , \ee
where the momentum fractions, $x_1$ and $x_2$, are
\be x_{1,2} = \frac{m_T}{\sqrt{s}} (e^{\pm y_Q} + e^{\pm y_{\overline Q}})
\, \, , \ee and $m_T = \sqrt{m_Q^2 + p_T^2}$.
The target fraction, $x_2$, decreases with rapidity while the
projectile fraction, $x_1$, increases.  Both increase with $p_T$.
The quark $p_T$ distribution is determined by
$d\widehat{\sigma}_{ij}/dt \propto 1/m_T^4$, as discussed in \cite{Ellis}.

At NLO, ${\cal O}(\alpha_s^3)$, in addition to real and virtual
corrections to the Born diagrams, quark-gluon scattering,
$q (\overline q) g \rightarrow Q \overline Q q (\overline q)$,
is also included.  At the Born level, this has been interpreted
as scattering a heavy quark in the hadron sea with a light quark or
gluon and referred to as flavor excitation\footnote{In deep-inelastic
scattering when $Q^2\gg m_Q^2$ the heavy
quark can be treated as massless
and absorbed into the parton densities but when $Q^2 \sim m_Q^2$, flavor
creation diagrams are dominant \cite{smtu}.} \cite{Ellis}.  The relative
importance of the excitation diagrams to hadroproduction is shown in Ref.\
\cite{Meng}.

The total short-distance cross section, $\widehat{\sigma}_{ij}$, is \be
\widehat{\sigma}_{ij}(s,m_Q,\mu_F,\mu_R) = \frac{\alpha_s^2(\mu_R)}{m_Q^2}
\left\{ f^0_{ij}(\rho)
+ 4\pi \alpha_s(\mu_R) \left[f^1_{ij}(\rho) + \overline
f^1_{ij}(\rho)\ln(\mu_F^2/m_Q^2) \right] + {\cal O}(\alpha_s^2) \right\}
\, \, , \ee where $\rho = 4m_Q^2/s$ and the functions $f_{ij}^n$ are
coefficients of the perturbative expansion.
The Born contribution is given by $f_{ij}^0$ and vanishes when $ij =
gq$, $g \overline q$.  The Born coefficients are always positive and
$f^0_{ij} \rightarrow 0$ as $\rho \rightarrow 0$.  The NLO coefficients
can be either positive or negative and tend to finite values as $\rho
\rightarrow 0$ in the $gg$ and $q \overline q$ channels.
The mass factorization terms, $\overline f^1_{ij}$,
only contribute when $\mu_R \neq m_Q$.  The $gg$ and $qg$ contributions are
dominant in the high energy limit due to $t$-channel gluon exchange.
The NLO $gg$ corrections are very large since
$f^1_{gg}/f^0_{gg} \rightarrow \infty$ as $\rho \rightarrow 0$,
indicating that a small $x$ resummation is needed.
For completeness, in Fig.\ 1, the coefficients are reproduced
from Ref.\ \cite{NDE1}
as a function of $1/\sqrt{\rho} = \sqrt{s}/2m_Q$.  Although the coefficients
are shown for $1/\sqrt{\rho}
< 200$ only, $f^1_{ij}$ and $\overline f^1_{ij}$ have reached a plateau.
The maximum value of $1/\sqrt{\rho}$ for a given energy is $1/\sqrt{\rho_{\rm
max}} = \sqrt{S}/2m_Q$, so that $3 \leq 1/\sqrt{\rho_{\rm max}}
\leq 12$ at typical fixed-target
energies while $1/\sqrt{\rho_{\rm max}} > 20$ for $b \overline b$
production and $>77$ for $c \overline c$ production at RHIC and LHC.

The physical cross section
should be independent of the scale: the dependence in
eq.\ (8) introduces an unphysical parameter.  If the
perturbative expansion is convergent, {\it i.e.}\ if further higher-order
corrections are small, at some scale the
${\cal O}(\alpha_s^{n+1})$ contribution to the cross section should be
smaller than the ${\cal O}(\alpha_s^{n})$ contribution\footnote{The
order of the expansion is represented by $n$ where $n\geq 2$ for $Q
\overline Q$ production.  A calculation to order ${\cal O}(\alpha_s^n)$
introduces corrections at the order ${\cal O}(\alpha_s^{n+1})$.  Thus the $\mu$
dependence should decrease when additional higher-order corrections are
included if the perturbation theory converges.}.  If the
scale dependence is strong, the perturbative expansion is untrustworthy
\cite{Ellis}.  Since $K_{\rm th}-1>1$,
further higher-order corrections are needed, particularly for charm and bottom
quarks which are rather ``light" when $\sqrt{S}$ is large.  Although the scales
are, in principle, independent, we take
$\mu_F = \mu_R = \mu$ unless otherwise noted because this assumption is
inherent in global analyses of parton densities.  Sometimes we use the notation
$\mu = \sqrt{Q^2}$.

The two loop value of the strong coupling constant, $\alpha_s$,
depends on the number of active flavors, $f$, and the appropriate
value of $\Lambda_{\rm QCD}$ for the number of flavors,
$\Lambda_f$,
\be \alpha_s(\mu,f) = \frac{1}{b_f \ln(\mu^2/\Lambda_f^2)} \left[ 1 -
\frac{b_f^\prime \ln \ln(\mu^2/\Lambda_f^2)}{b_f \ln(\mu^2/\Lambda_f^2)}
\right] \, \, , \ee
where $b_f = (33-2f)/12\pi$ and $b_f^\prime = (153-19f)/(2\pi(33-2f))$.
For charm and bottom production, $f=3$
and 4.  Parton densities are calculated assuming
the heavy quark
contributions turn on at $\mu = m_Q$ and
$\alpha_s(m_Q,f) = \alpha_s(m_Q,f+1)$.  Above this threshold, some groups, {\it
e.g.} \cite{D0}, treat the heavy quark as massless in the evolution of the
parton densities, introducing some overcounting into production calculations.
Also note that the heavy quark mass threshold chosen in the global analysis of
parton densities may differ from the mass used in calculations.

We have used two sets of recent parton distribution
functions\footnote{All available parton
distribution functions are contained in the package PDFLIB \cite{PDF},
available in the CERN library routines.}, GRV HO \cite{GRV} and
MRS D$-^\prime$
\cite{D0}.  GRV HO has a low initial scale, $Q_{0,{\rm GRV}}^2 =
0.3$ GeV$^2$, with
valence-like parton distributions, therefore evolving very quickly with $Q^2$.
MRS D$-^\prime$ has a more conventional initial scale, $Q_{0,{\rm MRS}}^2 =
5$ GeV$^2$, and
sea quark and gluon densities
that grow as $\sim x^{-1/2}$ when $x \rightarrow 0$.  The difference in $Q_0^2$
produces the contrast in $Q \overline Q$ production between the two
sets.  Both are compatible
with the recent deep-inelastic scattering data from
HERA \cite{HERA} although both are on the high side of the data.
We also use the SMRS P2 \cite{SMRS} and GRV HO pion
\cite{GRVpi} densities.  The GRV HO pion distributions
are obtained from their proton set and have a similar behavior at low $x$.
The SMRS P2 distributions are based on an older set of
nucleon distributions with a different value of $\Lambda_4$
than MRS D$-^\prime$.  In this case, we evaluate $\alpha_s$ at the MRS
D$-^\prime$ value of $\Lambda_4$.  Also, note that at $Q_{0, \rm MRS}^2$,
the pion sea quark and gluon
distributions are assumed to be constant as $x \rightarrow 0$, incompatible
with the MRS D$-^\prime$ low $x$ behavior.

Since we will discuss both $p(\overline p)p$ and $\pi^- p$ production,
it is instructive to show the quark and gluon momentum distributions for
protons and pions.  In Fig.\ 2(a) and (c) we show $xf(x) = x(u_V(x) +
d_V(x) + 2(\overline u(x) + \overline d(x) + \overline s(x))) = x(u_V(x) +
d_V(x) + S(x))$ and $xg(x)$ at $Q^2 = 5$, 25,
and 100 GeV$^2$.  The lower curve for each set
corresponds to the lowest $Q^2$, typical for charm
production.  The middle curve, at 25 GeV$^2$, represents charm
production at $p_T \simeq 3.5 m_c$ and $b$ production at $\mu \simeq m_b$.
The highest $Q^2$ is equivalent to $c$ and $b$ production at $p_T > m_Q$.
In 2(b) and (d), we show the parton distributions as a function of $Q^2$
for $x = 0.0025$, 0.007, and 0.15.  Here the lowest $x$ values
correspond to $b$ production at CDF and UA1.
We choose $x = 0.15$ as typical of central
charm production at fixed target energies.
The curves are MRS D$-^\prime$ (solid), GRV HO (dashed), SMRS P2 (dot-dashed),
and GRV HO pion (dotted).  At fixed $Q^2$, parton
distributions that initially increase as $x \rightarrow 0$ are depleted at
higher $x$ compared to
constant initial distributions such as SMRS P2.
The pion quark distributions are larger than the proton distributions
as $x \rightarrow 1$.
When the partonic cross sections are convoluted with the parton
densities, $gg$ and $qg$ processes dominate because the flux, $\propto
F_i^A(x_1,\mu_F) F_j^B(x_2,\mu_F)$\footnote{In this notation, {\it e.g.}\
$F_g(x) = g(x)$.}, is large.  In fact, although the gluon
density is larger at small $x$, as $x$ increases, it drops
faster than the quark density so that, at some $x_1 x_2$, $F_q^A F_g^B >
F_g^A F_g^B$.  Eventually, as $x_1 x_2 \rightarrow 1$, $q \overline
q$ annihilation dominates since the valence quark density is most important at
large $x$.

In Fig.\ 3(a), we show the scale dependence of the NLO
calculations
with the $pp$ and $pA$ data on $\sigma_{c \overline c}^{\rm
tot} (S)$ \cite{Reu,Appel,MLM1}
assuming a linear nuclear dependence \cite{alv}.
The scale is varied between $m_Q/2$ and $2m_Q$ with $1.2 < m_Q < 1.8$ GeV
to show the range of theoretical uncertainty \cite{NDE1,MLM1}.
We further assumed that $\sigma(D_s) \approx 0.2 \, \sigma(D^0 + D^+)$ and
$\sigma(\Lambda_c) \simeq 0.3 \, \sigma(D^0 + D^+)$ so that $\sigma_{c
\overline
c}^{\rm tot} \approx 1.5 \sigma(D \overline D)$, as in Ref.\ \cite{MLM1}.
Since $m_c <
Q_{0,{\rm MRS}}$, we take $\mu_F=2m_c$ for the calculations with the MRS
parton densities.  The three solid curves are calculated with MRS D$-^\prime$
densities and $m_c = 1.2$ GeV, $\mu_R = m_c/2$ (upper); $m_c = 1.2$ GeV,
$\mu_R = 2m_c$ (middle); and $m_c = 1.8$ GeV, $\mu_R = 2m_c$ (lower).  The
difference between the upper and lower curves gives the maximum variation
of $\sigma_{c \overline c}^{\rm tot}$ for these densities, a factor of 90 for
$\sqrt{S} = 20$ GeV and 20 for $\sqrt{S} = 14$ TeV.
If the MRS D$0^\prime$ set is used at the same mass and scale, the cross
section is slightly larger at fixed-target energies because the D$0^\prime$
gluon density is larger for $x > 0.06$ but since $xg(x
\rightarrow 0,Q^2_{0,{\rm MRS}})
\rightarrow {\rm constant}$, $\sigma_{c \overline c}^{\rm
tot}({\rm D}-^\prime)/\sigma_{c \overline c}^{\rm
tot}({\rm D}0^\prime) \approx 20$ at 14 TeV.  The corresponding
$K$ factors are given by the solid line in Fig.\ 3(b)--$K_{\rm th}$
has very little scale dependence with these parton densities.

Because the GRV HO distributions have a much lower $Q_0$, we take
$\mu_R=\mu_F$.  The dot-dashed and dotted curves show $\mu = m_c/2$
and $\mu = 2m_c$.  The upper curves have $m_c = 1.2$
GeV, the lower, $m_c = 1.8$ GeV.  Even though the cross section is larger
at low $\sqrt{S}$ for $\mu = m_c/2$, as the energy increases, the small $x$
gluon density at the higher scale becomes the dominant feature, causing the
crossover shown at high energy.  In fact, because the gluon
density starts out valence-like at $Q_{0,{\rm GRV}}$, it is
almost constant for $x$ values probed at $\sqrt{S} > 100$ GeV
for $\mu = 0.6$ GeV,
causing the sudden flattening of the upper dot-dashed curve.  The variation in
$K_{\rm th}$ is large when $\mu =
m_c/2$.  When $\mu = 2m_c$, $K_{\rm th}$ is approximately the same as the
MRS distributions.  However, the maximum variation
of $\sigma_{c \overline c}^{\rm tot}$ is smaller for the GRV HO densities,
a factor of 60 for
$\sqrt{S} = 20$ GeV and 6 for $\sqrt{S} = 14$ TeV (excluding $\mu = 0.6$ GeV).

Previously \cite{HPC}, the NLO calculations were compared to the
data to  fix $m_c$ and $\mu$ at $K_{\rm exp}^{\rm NLO} \sim 1$ to provide
an estimate that could
be extrapolated to nuclear collider energies.  Reasonable
agreement was found for $m_c = 1.2$ GeV, $\mu = 2m_c$ for MRS D$-^\prime$
(central solid curve) and $m_c = 1.3$ GeV, $\mu = m_c$ for GRV HO (dashed
curve)\footnote{A comparison to the $c \overline c$ data with $\pi^-$ beams
\cite{Reu,Appel,MLM1}
using the same parameters gives agreement with the data at a similar
level.}.  Note however that both curves tend to underestimate
$\sigma_{c \overline c}^{\rm tot}$ with $K_{\rm exp}^{\rm NLO}
\sim 1.1 - 2$.  In the
range of the parameter space defined by $m_Q$, $\mu_R$ and $\mu_F$,
$K_{\rm exp}^{\rm NLO}$ can be reduced to unity.
However, it is questionable if
the mass and scale values needed for $K_{\rm exp}^{\rm NLO} \sim 1$
are consistent with a perturbative treatment
and with the defined limits of the parton density
distributions\footnote{Recently, $m_c =1.5$ GeV  was found
to be compatible with this data with some essential caveats:
$\mu_F$ and $\mu_R$ were varied independently and out-of-date
parton distributions fit with several values of
$\Lambda_{\rm QCD}$ were used \cite{MLM1}.
Decreasing $\mu_R$ with respect to $\mu_F$ and
increasing $\Lambda_{\rm QCD}$ result in significantly larger cross
sections for a given $m_c$.  Additionally, different parton densities were used
in the calculations of $\sigma_{Q \overline Q}^{\rm tot}$ and high energy $b$
production.}.  It is also not clear that the NNLO corrections to heavy quark
production would
not be at least as large as the NLO corrections, particularly
when $m_Q \ll \sqrt{S}$, even though for high-mass Drell-Yan production at NNLO
$\sigma_{\rm NNLO}/\sigma_{\rm NLO} \sim
1.1$-1.3 \cite{DYK}, due to cancellations among the different channels.

However, the variation in $\sigma_{b \overline b}^{\rm tot}$
is much smaller than $\sigma_{c \overline c}^{\rm tot}$.
For $\mu = m_b$, $\sigma(m_b=4.5 \, {\rm GeV})/\sigma(m_b=5 \, {\rm GeV})
\approx 25$ at $\sqrt{S} = 40$ GeV and 1.4 for 14 TeV
while for $m_b=4.75$ GeV,
$\sigma(\mu_b = 0.5m_b)/\sigma(\mu_b = 2m_b) \approx 3$ at $\sqrt{S} = 40$ GeV
and 1.2 at 14 TeV.  The $K$ factor is also smaller; $K_{\rm th}
\approx 1.2-2.5$.

In Fig.\ 4 we show the scale
variation of $\sigma_{c \overline c}^{\rm tot}$ for $\pi^- p$
production at 340 GeV (a) and $pp$ production at 800 GeV (b) using
the GRV HO pion and nucleon distributions.
As $\mu/m_c$ increases at fixed energy, the cross section
varies less rapidly but $K_{\rm th}$ increases for $\mu > m_c$.
There is no value of $\mu$ where the NLO corrections to the Born cross section
are minimal, {\it i.e.}\ no optimal scale \cite{PMS}.

Although there is no physical reason for assuming that $\mu_F$ and $\mu_R$
are different because they are not separated in analyses of the parton
densities, we also show the change of the $c \overline c$ cross
section in $\pi^- p$
production with $p_\pi = 500$ GeV induced by varying the scales independently:
$\mu_F = m_c$ in 4(c) and $\mu_R=m_c$ in 4(d).  In addition
to the GRV HO results, we include the calculations with the MRS D$-^\prime$
parton distributions and $\mu_F = 2 m_c$ in (c) and
$\mu_R = 2m_c$ in (d) at NLO (dot-dashed) and the Born level (dotted).
The running of $\alpha_s$ is the source
of the strong dependence observed in (c).  When $\mu_R < m_c$, the
coefficients $\overline f_{ij}^1$, usually negative, produce an overall
enhancement of $\widehat{\sigma}_{ij}$.  However for $\mu_R > m_c$, the
variation of $\widehat{\sigma}_{ij}$
with $\mu$ decreases since the NLO corrections partially
cancel each other.  There is no strong effect on $K_{\rm th}$.
In (d), the increase in $\sigma_{c \overline c}^{\rm tot}$ for $\mu_F<m_c$
with the GRV HO distributions occurs because at $\mu_F \approx
Q_{0,{\rm GRV}}$ and low $x$, the sea quark and gluon
distributions are valence-like \cite{GRV}.  The MRS D$-^\prime$ variation
is negligible since
$\mu_F < Q_{0,{\rm MRS}}$ for $\mu_F/m_c < 2.7$, thus
$\mu_F \equiv Q_{0,{\rm MRS}}$.

Until now, we have discussed $\sigma_{Q \overline Q}^{\rm tot}$ where
$\mu \propto m_Q$.  However, for
single inclusive or double differential distributions, it may be more
appropriate to choose a scale proportional to the transverse momentum
of the heavy
quark, $p_T$, or its transverse mass, $m_T$.
The transverse momentum is a natural scale in jet and prompt photon production
with massless final-state particles. Additionally, since $p_T$, $m_T \propto
(t-m_Q^2)(u-m_Q^2)/s \rightarrow (T -m_Q^2)(U-m_Q^2)/S$, this choice is
invariant under scale transformations of the initial state momenta and results
in a simple relation between the partonic and hadronic variables, unlike
the choice $\mu = s$.  A constant scale
would be appropriate if $m_Q \simeq p_T$, but generally $m_c \ll p_T$ and
$m_b \ll p_T$ at collider energies. Therefore the $p_T$ dependent scale
absorbs (resums) large logarithmic terms such as
$\ln(p_T/m_Q)$ appearing when $p_T \gg m_Q$ and producing collinear
divergences \cite{NDE2} which are unregulated if $\mu = m_Q$.
We examine the $p_T$ distributions of
$b$ quark production at collider energies for two specific cases, a
constant scale, $\mu \propto m_b$, and running scale,
$\mu \propto m_T$, to see if the data favors a particular scale.
We also show the $p_T$ distributions and $K_{\rm th}(p_T)$ for charmed
quarks at fixed target energies.

We first compare the constant and running scales for single $b$
production from  the $p \overline p$
colliders at $\sqrt{S} = 630$ GeV and 1.8 TeV.  The measurements are
integrated over $p_T$ above each $p_{T,{\rm min}}$.  We use $m_b = 4.75$ GeV.
The NLO calculations with the GRV HO distributions
are shown in Fig.\ 5(a) for UA1 \cite{UA1} and Fig.\ 5(b) for CDF
\cite{CDF2} and D0 \cite{DZ}. The solid curves show $\mu = m_T$,
the dashed curves, $\mu = m_b$.  Neither are clearly favored
by the data.  The
Tevatron data is also compared to calculations with $\mu =
m_T/4$ and $m_b/4$.  However $\mu = m_b/4$ is in clear
contradiction with the data.

In Fig.\ 6(a) we show the charmed quark $p_T$ distributions for $\pi^-
p$ production with $p_{\rm lab} = 500$ GeV using the GRV HO
distributions.
The solid and dashed curves are the NLO and Born results for $\mu =
m_T$ while the dot-dashed and dotted curves are the calculations
with $\mu = m_c$.  The NLO result drops below
the Born for $p_T > 4$ GeV.  High statistics charm measurements
should be able to detect a decrease in the
slope of the $p_T$ distribution
for $p_T > 4$ GeV if nature favors the constant scale.  No such change has been
observed at present although data in this region is rather scarce.
The resulting ratios $K_{\rm th}(p_T)$
are shown in Fig.\ 6(b) for the running scale (solid curve)
and the fixed scale (dashed curve).  When the scale increases with $p_T$,
$K_{\rm th}$ is relatively constant or decreases slightly.  However, the fixed
scale produces a strong decrease of $K_{\rm th}$ with $p_T$.

The calculations in Fig.\ 6(a) and (b) are particularly illustrative of
both how the production processes compete with each other and the relative
importance of the NLO corrections for each process.  The crossover of the
Born and NLO $p_T$ distributions with the fixed scale, occuring at $p_T
\approx 4$ GeV,
can be understood by an examination of the coefficients of the perturbative
expansion.
To schematically illustrate the $p_T$ dependence, we can
take $1/\sqrt{\rho} \approx \sqrt{s}/2m_T$ in Fig.\ 1.
In the region $0 < p_T < 5$ GeV, $1/\sqrt{\rho} < 12$.  For $1/\sqrt{\rho} \leq
4$, $f^0_{gg} > f^1_{gg}$ while $f^1_{gg}$ dominates for $1/\sqrt{\rho} \geq
4$.  Both coefficients
are always positive although $\overline f^1_{gg}$ is negative
for $1/\sqrt{\rho} > 2.2$.  The
coefficient $f^1_{qg}$ is positive when $1/\sqrt{\rho} > 2.2$
although smaller than
$f^1_{gg}$; $\overline f^1_{qg}$ is negative and $f^1_{qg} > |\overline
f^1_{qg} |$ for most values of $1/\sqrt{\rho}$.
The NLO correction to $q \overline q$ annihilation, $f^1_{q
\overline q}$, is negative for $1.4 < 1/\sqrt{\rho} < 4.5$.
When $\mu \neq m_Q$,
$\overline f^1_{q \overline q}$ can partially compensate for this negative
contribution.  As $p_T$
increases, $1/\sqrt{\rho}$ enters the region of negative $f^1_{q
\overline q}$ at
$p_T \approx 4$ GeV while $f^0_{q \overline q}$ is large and positive.
At the same time the Born contribution to $gg$ fusion becomes larger than the
NLO correction.  These competing terms produce the crossover observed in Fig.\
6(a) for the constant scale.  The same effect causes the turnover in the
dashed curve of Fig.\ 5(a) for $p_{T, {\rm min}} \approx 50$ GeV.  Thus the
constant scale, although introducing divergences,
shows the structure of the NLO corrections.

Assuming that $\mu \propto m_T$ softens these effects, partially because of
the $\overline f^1_{ij}$ contributions but mainly due to the evolution of the
parton densities.  As $p_T$ increases, $x_1$ and $x_2$ also increase
so that while the $Q^2$ evolution increases the parton density at fixed $x$,
the higher $x$ values actually deplete the parton flux.  This relative balance
produces a nearly constant $K_{\rm th}$.  Results with the MRS
distributions are shown in Fig.\ 6(c) and (d).  Since $\mu = 2m_c$ for the
fixed scale, a larger
$p_T$ is needed to enter the crossover region in $1/\sqrt{\rho}$.

We now turn to the scale dependence of heavy quark distributions at nuclear
collider energies.  While the NLO calculations are needed for the $p_T$
dependence
of $Q \overline Q$ pair production and decay, trivial at the Born level,
it would be convenient if the
other relevant distributions could be modeled by
the Born distributions to within a constant $K$ factor.
The $Q \overline Q$ pair distributions are essential to determine the heavy
quark contribution to lepton pair production,
In Figs.\ 7-10 we show $K_{\rm th}$ for $\sqrt{s} = 200$ GeV (RHIC) and
5.5 TeV (LHC) for $c$
and $b$ production assuming $\mu = m_T$ for single inclusive quark $x_F$,
rapidity, and $p_T^2$
distributions in (a), (c), and (e) respectively.  The $Q \overline Q$ pair
$x_F$, $y$, and invariant mass distributions are shown in (b), (d), and (f).
For the pair distributions, we assume
$\mu \propto \sqrt{m_Q^2 + (p_{T, Q}^2 + p_{T, \overline Q}^2)/2}$.
The results with the GRV HO parton densities are given by the circles while
the diamonds are the MRS D$-^\prime$ results. (For the NLO distributions
themselves, see Ref.\ \cite{HPC}.)  With the running scale,
$K_{\rm th}$ is indeed nearly constant for these quantities, even in
a regime where $m_Q \ll \sqrt{S}$.  Some variations occur
near the boundaries of phase space, seen in the $c$ and $b$ quark rapidity
distributions at $\sqrt{S} = 200$ GeV.  The $K$ factors are independent of
parton density except for charm production at $\sqrt{S} = 5.5$ TeV where
$K_{\rm th}$ is larger for the GRV HO distributions.  For the charmed quark,
$K_{\rm
th}(p_T^2)$ increases 50\% at $\sqrt{S} = 200$ GeV and a factor of
two at $\sqrt{S} = 5.5$ TeV.  These increases perhaps indicate the
appearance of large logarithms that need to be resummed in the $p_T$
distributions.  Although some variation appears, event generators for heavy
quark production can, with relative confidence,
scale all non-trivial Born results\footnote{A constant scale is assumed in
\cite{Ina}, leading to the conclusion that $K_{\rm th}(p_T)$ is a strongly
increasing
function of $p_T$. Our calculations with the constant scale confirm these
results in the $p_T$ region explored in \cite{Ina}.  However, we note that for
$p_T > 6$-8 GeV at RHIC $K_{\rm th}$ decreases again while at the LHC $K_{\rm
th}$ reaches a plateau when $p_T \sim 10$ GeV.}.

Thus for heavy quark and quark pair distributions calculable
at the Born level, $K_{\rm th}$ is nearly constant provided that $\mu \propto
m_T$.
The actual value of $K_{\rm th}$ can be determined by a comparison of the NLO
and Born total cross sections.  We have set the scale by
making $K_{\rm exp}^{\rm NLO}$ close to unity to better estimate production at
nuclear colliders.  Other, more
sophisticated methods can be used to optimize the scale \cite{PMS} or
relate it to other scales in perturbative QCD \cite{BLM}.
Further theoretical advances can perhaps also reduce the uncertainties.
A resummation of the soft gluon contribution to top production near threshold
has been performed \cite{Jack}.  This near-threshold resummation may also be
applied to charm and bottom \cite{KS} production at fixed-target energies,
suggesting a more appropriate scale.

An understanding of heavy quark production is essential to correctly interpret
data from RHIC and the LHC.  At these nuclear colliders, $Q \overline Q$
production is in the high-energy limit where large logarithmic
corrections to $\widehat{\sigma}_{ij}(\rho)$ are important.
A first attempt has been made to resum the small $x$ corrections to the heavy
quark production cross section in the $\rho \rightarrow 0$ regime \cite{EC}
which could lead to an improvement in the RHIC and LHC estimates.

We thank J.A. Appel, S. Gavin, B.W. Harris, P. Karchin, I. Sarcevic,
and J. Smith
for stimulating discussions and M.L. Mangano for help with the program package.
We also thank the BNL theory group for hospitality during part of this work.

\newpage

\begin{center}
{\bf Figure Captions}
\end{center}
\vspace{0.14in}
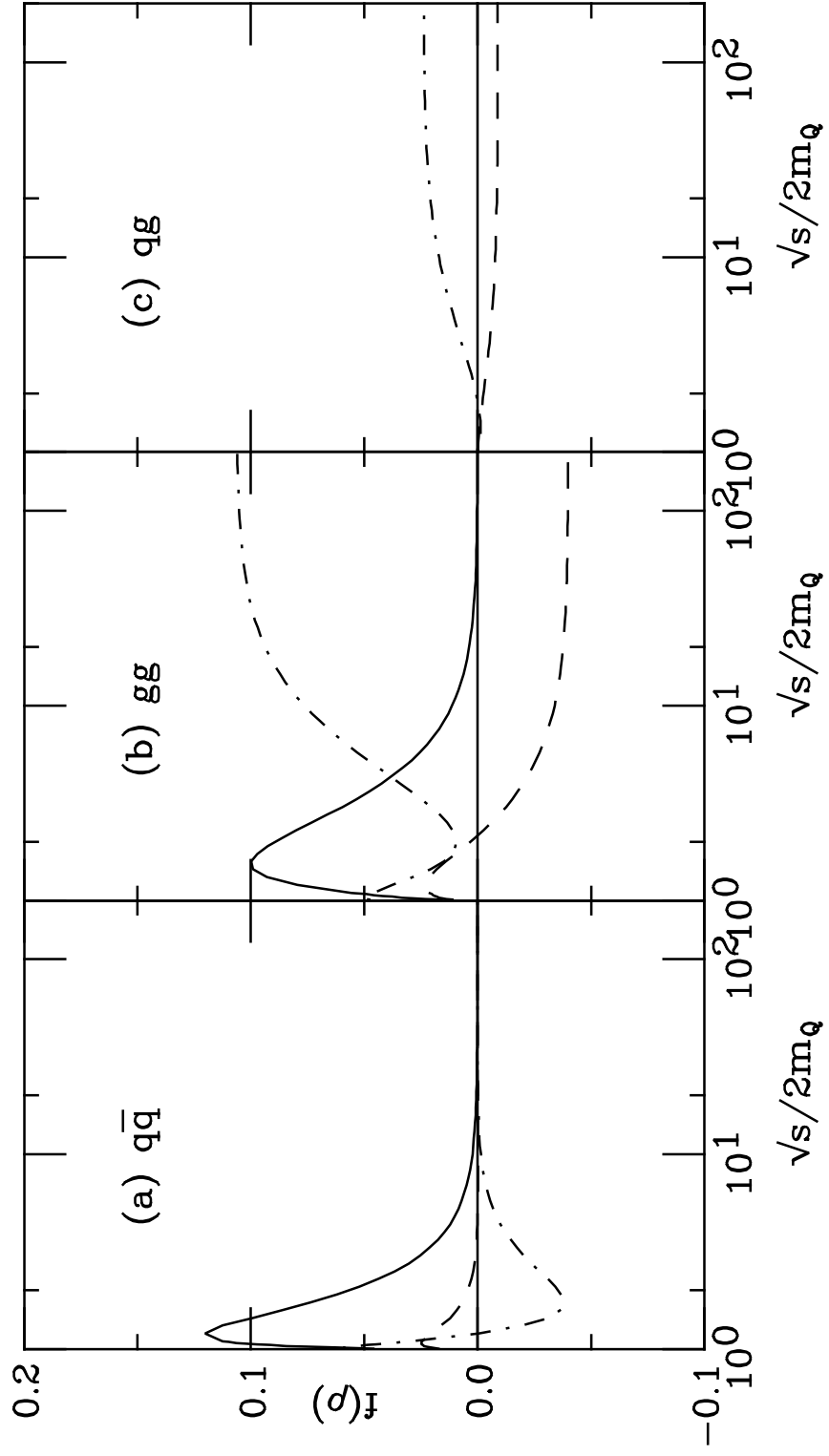
\noindent Figure 1. The coefficients of the perturbative expansion, $f(\rho)$,
given in Ref.\ \cite{NDE1} and shown as a function of $1/\sqrt{\rho}$.
The solid line is the Born contribution,
$f^0_{ij}$, the dot-dashed and dashed, the NLO corrections $f^1_{ij}$ and
$\overline f^1_{ij}$ for (a) $q \overline q$ annihilation, (b)
gluon fusion, and (c) quark-gluon scattering.

\vspace{0.14in}
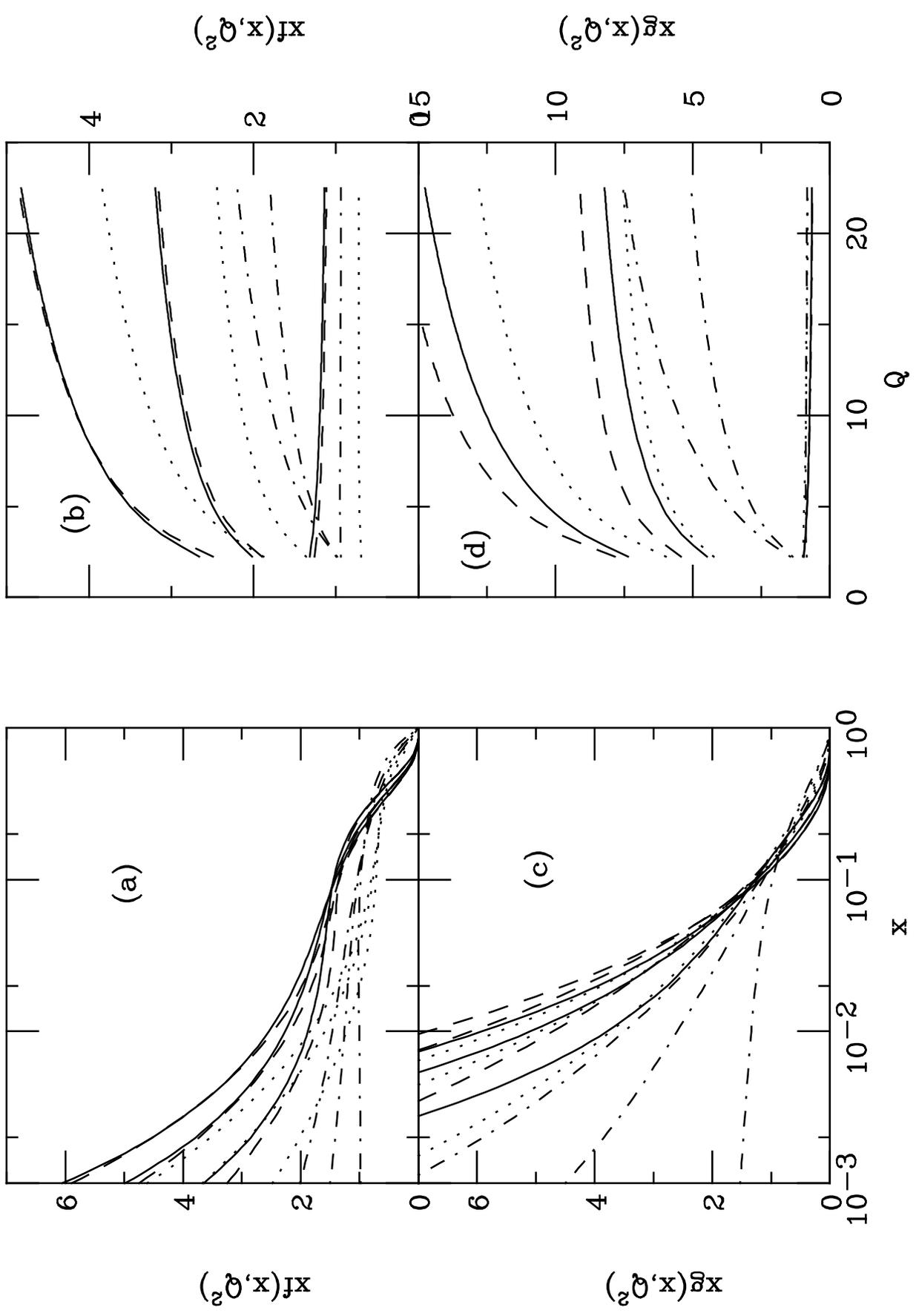
\noindent Figure 2.  The parton distributions $xf(x) = x(u_V(x) + d_V(x) +
S(x))$ and $xg(x)$ as a function of $x$ and $Q^2$.  In (a) and (c),
$Q^2 = 5$, 25, and 100 GeV$^2$ for the lower, middle, and upper curves
respectively. In (b) and (d), $x = 0.0025$, 0.007, and 0.15 in the upper,
middle, and lower curves respectively.
The curves are MRS D$-^\prime$ (solid), GRV HO (dashed), SMRS P2 (dot
dashed), and GRV HO pion (dotted).

\vspace{0.14in}
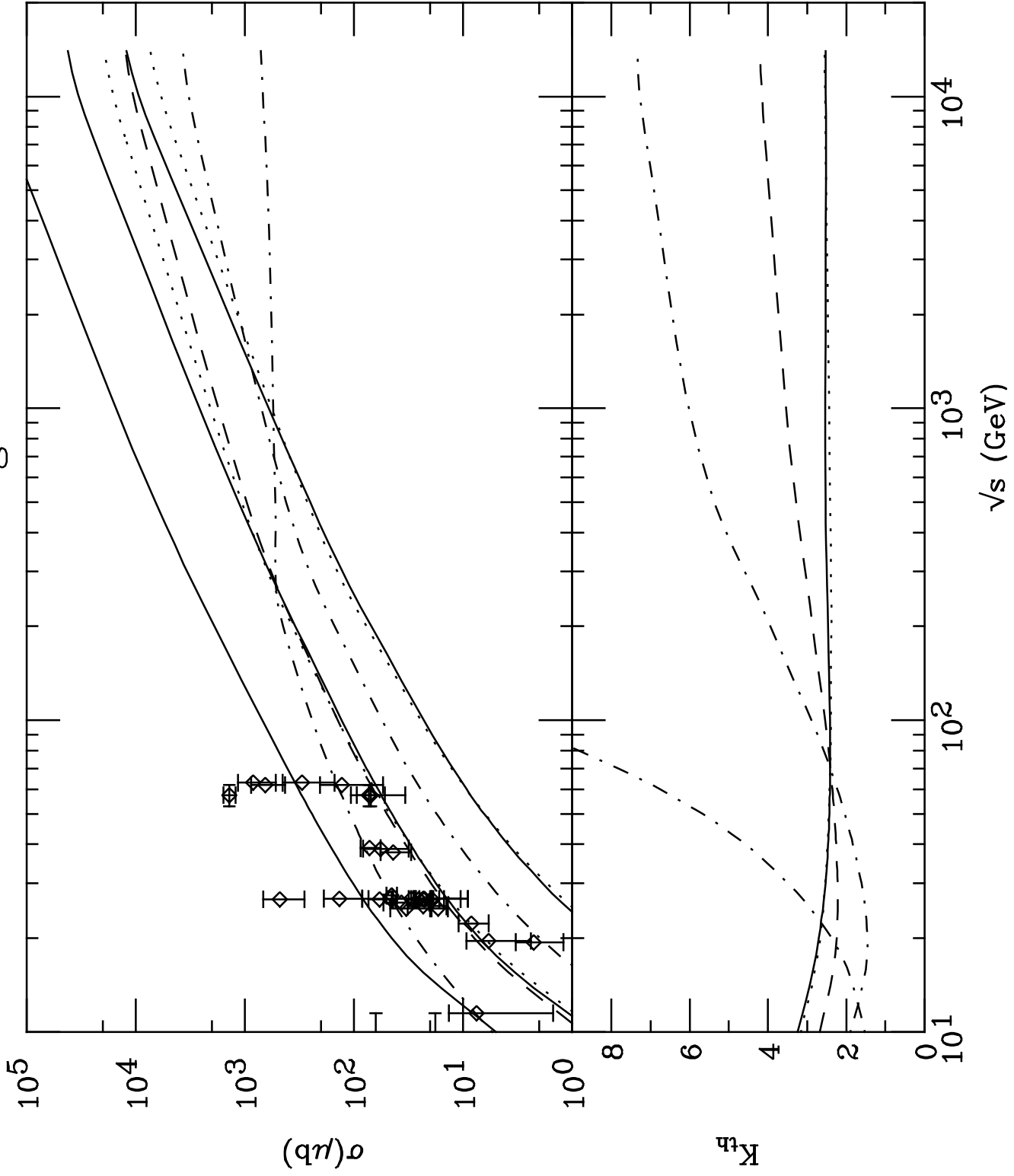
\noindent Figure 3.  The variation in $\sigma_{c \overline c}^{\rm tot}(S)$ and
$K_{\rm th}(S)$ with parton density, $m_c$ and $\mu$.
In (a) the three solid curves are calculated with MRS D$-^\prime$
densities and $m_c = 1.2$ GeV, $\mu_R = m_c/2$ (upper); $m_c = 1.2$ GeV,
$\mu_R = 2m_c$ (middle); and $m_c = 1.8$ GeV, $\mu_R = 2m_c$ (lower).
The other calculations are with the GRV HO densities.
The dot-dashed and dotted curves show $\mu = m_c/2$
and $\mu = 2m_c$.  The upper set has $m_c = 1.2$
GeV, the lower, $m_c = 1.8$ GeV.  The dashed curve is $m_c = 1.3$ GeV, $\mu =
m_c$. $K_{\rm th}$ is shown in (b).
The solid curve is MRS D$-^\prime$, $m_c
= 1.2$ GeV, $\mu = 2m_c$, the dotted curve GRV HO, $m_c = 1.2$ GeV, $\mu =
2m_c$.  The dot-dashed curves are GRV HO, $\mu = 0.5m_c$, and $m_c = 1.2$
GeV (upper), 1.8 GeV (lower).

\vspace{0.14in}
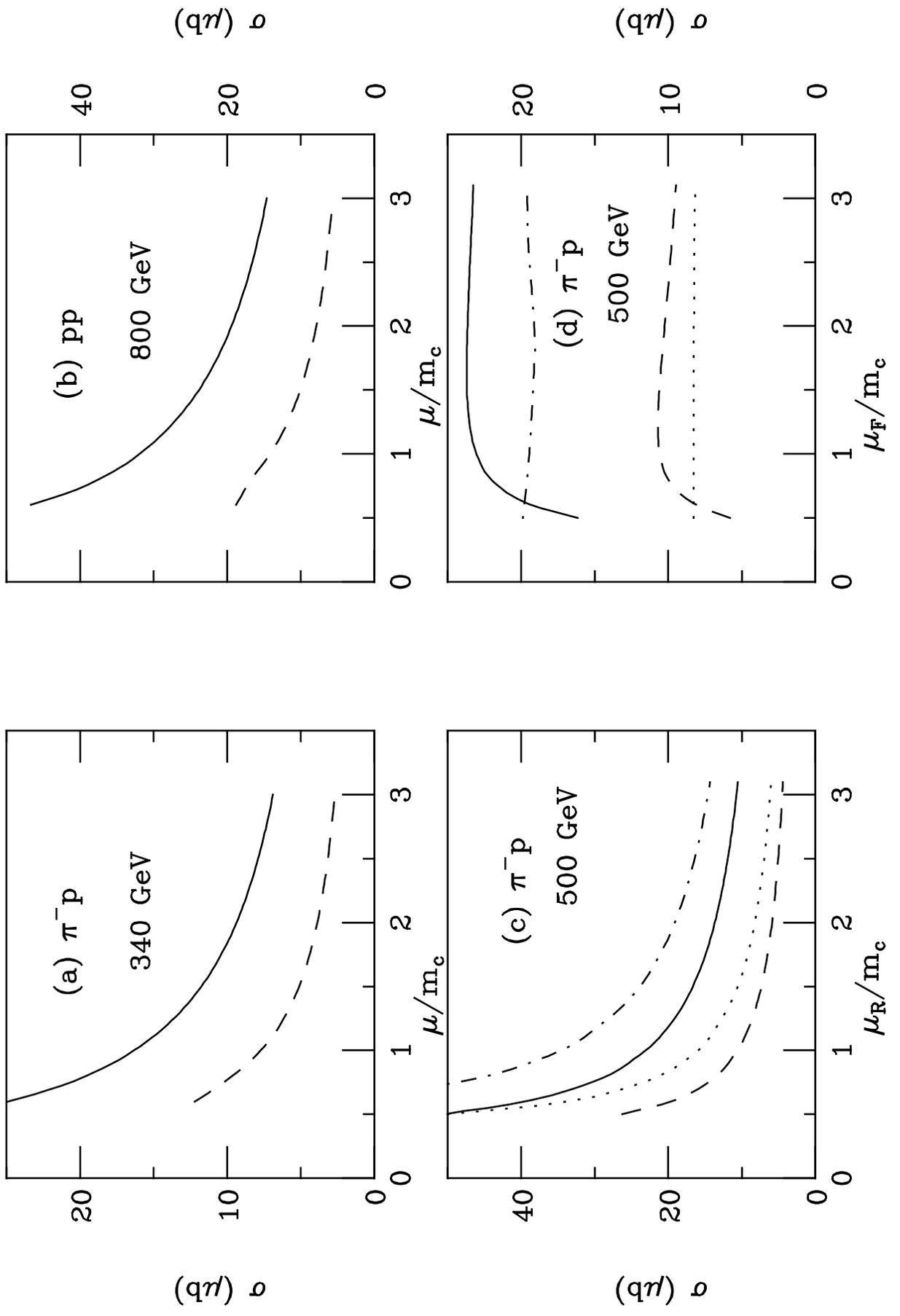
\noindent Figure 4.  The variation of $\sigma_{c \overline c}^{\rm tot}$
with $\mu = \mu_F = \mu_R$ for $\pi^- p$
production at 340 GeV (a) and $pp$ production at 800 GeV (b) using GRV HO.  The
change of the $c \overline c$ cross section in $\pi^- p$
production with $p_\pi = 500$ GeV from varying the scales
independently is shown for $\mu_F = nm_Q$ in (c) and for $\mu_R=nm_Q$ in (d)
where $n=1$ for GRV HO and 2 for MRS D$-^\prime$.  The curves are GRV HO
NLO (solid), Born (dashed), MRS D$-^\prime$ NLO (dot-dashed), Born (dotted).

\vspace{0.14in}
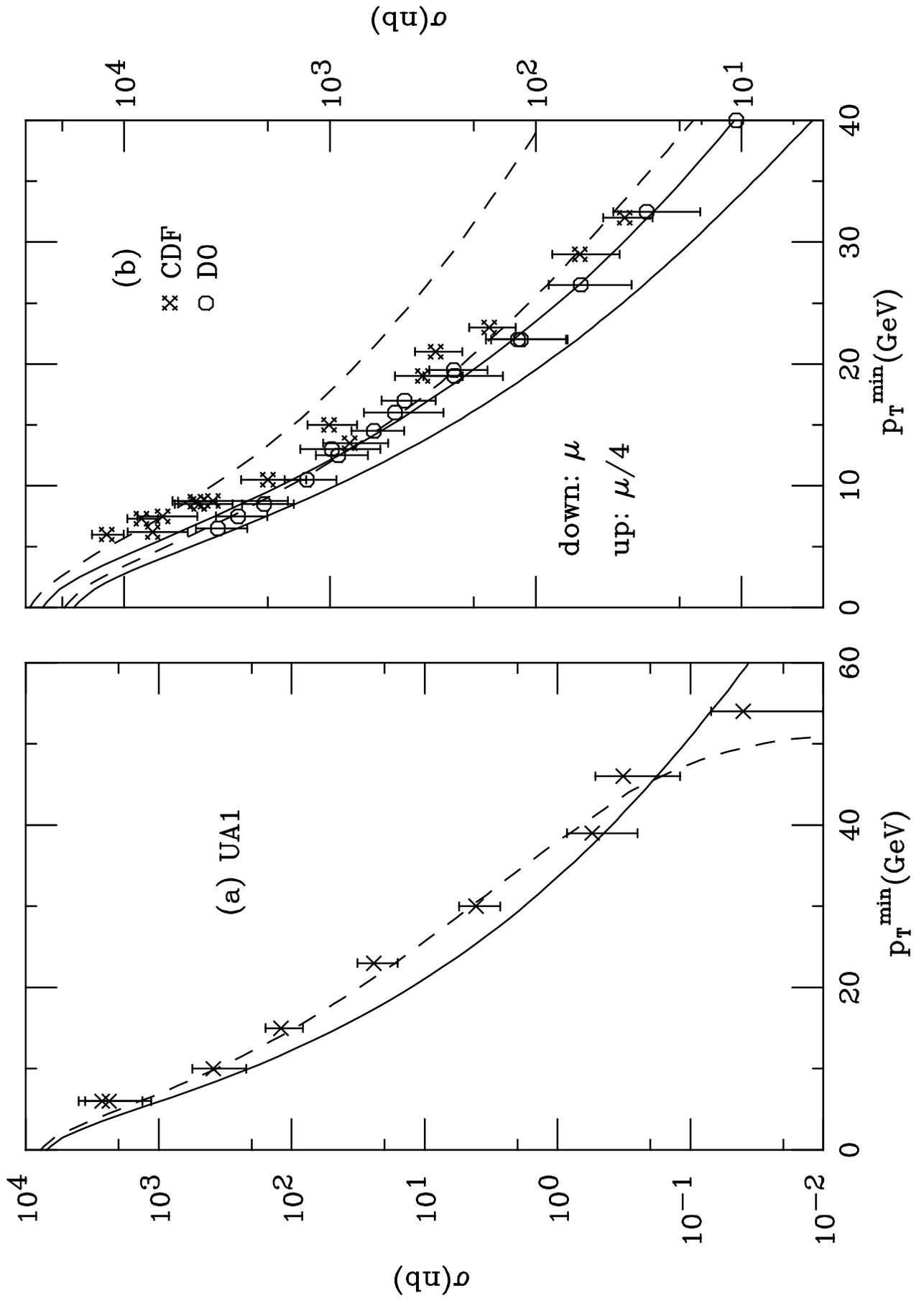
\noindent Figure 5.  Single $b$
production at NLO is compared with data from UA 1 (a) and CDF (b).
The solid curves show $\mu = m_T$ and the dashed curves, $\mu = m_b$.  We also
present $\mu = m_T/4$ and $m_b/4$ in the upper solid and dashed curves of (b).

\vspace{0.14in}
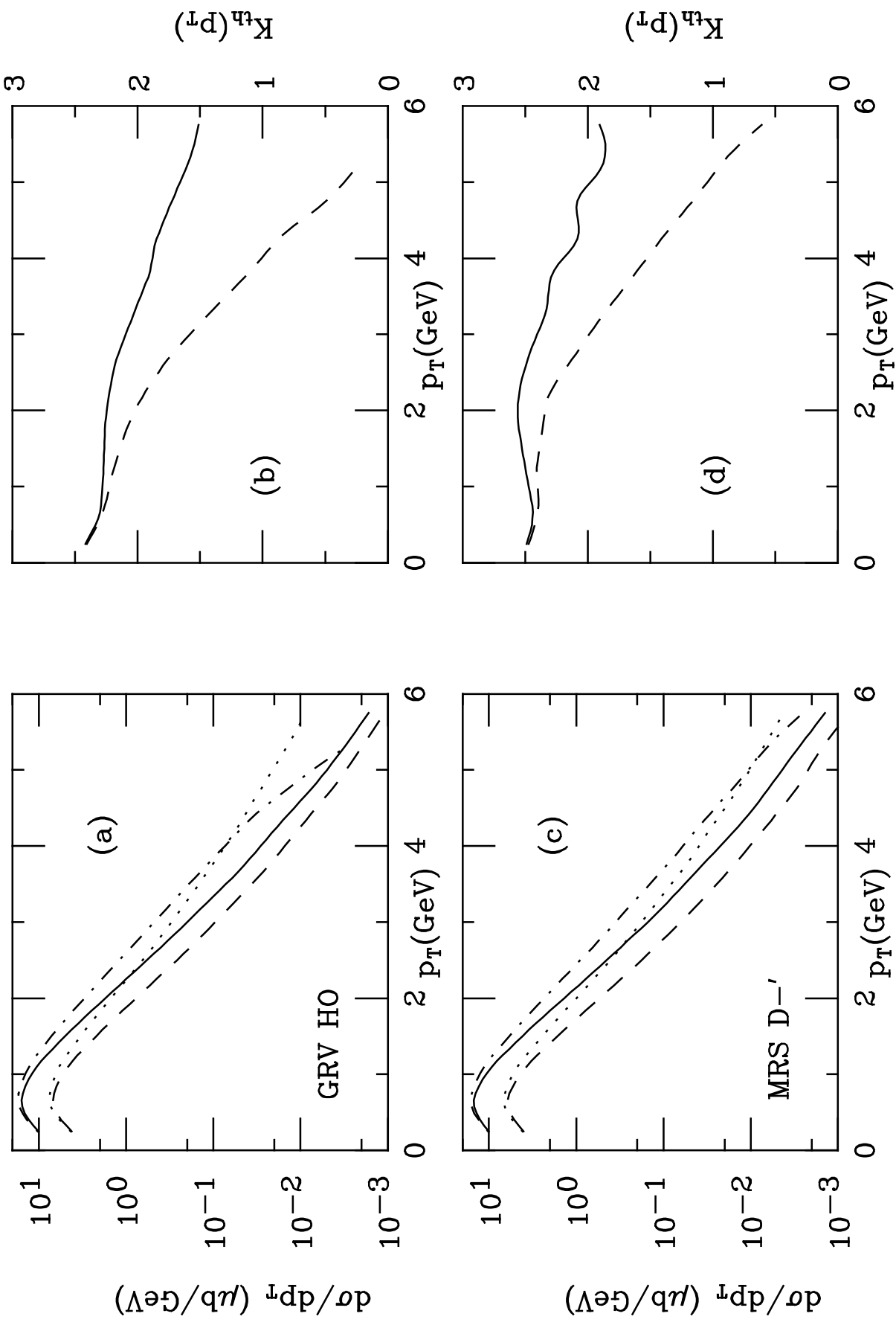
\noindent Figure 6. The charmed quark $p_T$ distribution for $\pi^-
p$ production with $p_{\rm lab} = 500$ GeV is shown in (a) for the GRV HO
parton distributions.
The solid and dashed curves are the NLO and Born results for $\mu =
m_T$ while the dot dashed and dotted curves are for $\mu = m_c$.
In (b) we show the $K_{\rm th}$ factors for the running scale (solid curve)
and the fixed scale (dashed curve).  The results with MRS D$-^\prime$
distributions are given in (c) and (d).

\vspace{0.14in}
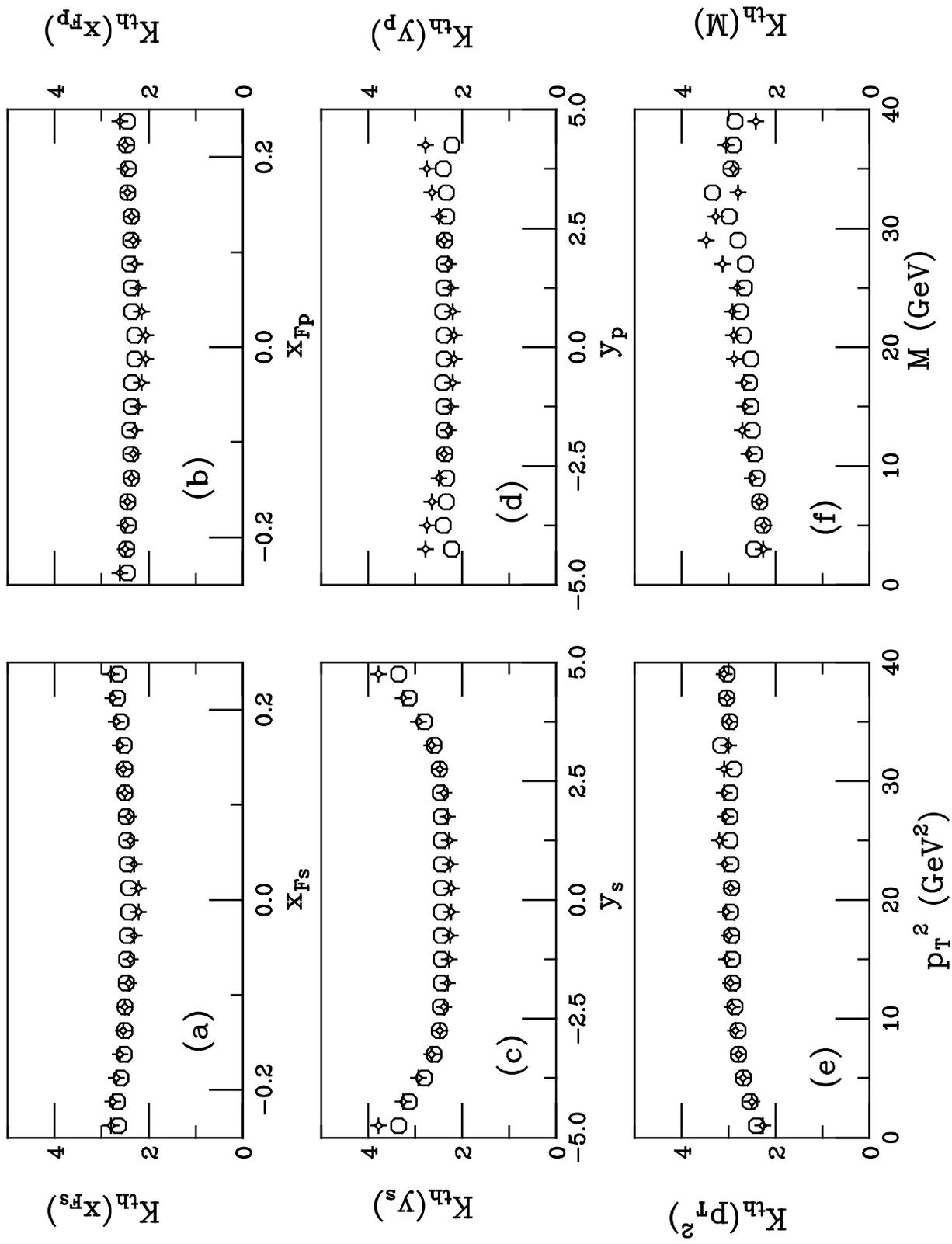
\noindent Figure 7.  The ratios $K_{\rm th}$ for charm production
at $\sqrt{S} = 200$ GeV assuming $\mu = m_T$ for single inclusive quark $x_F$,
rapidity, and $p_T^2$
distributions in (a), (c), and (e) respectively.  The $Q \overline Q$ pair
$x_F$, $y$, and invariant mass distributions are shown in (b), (d), and (f).
The GRV HO results are given by the circles, MRS D$-^\prime$,
the diamonds.

\vspace{0.14in}
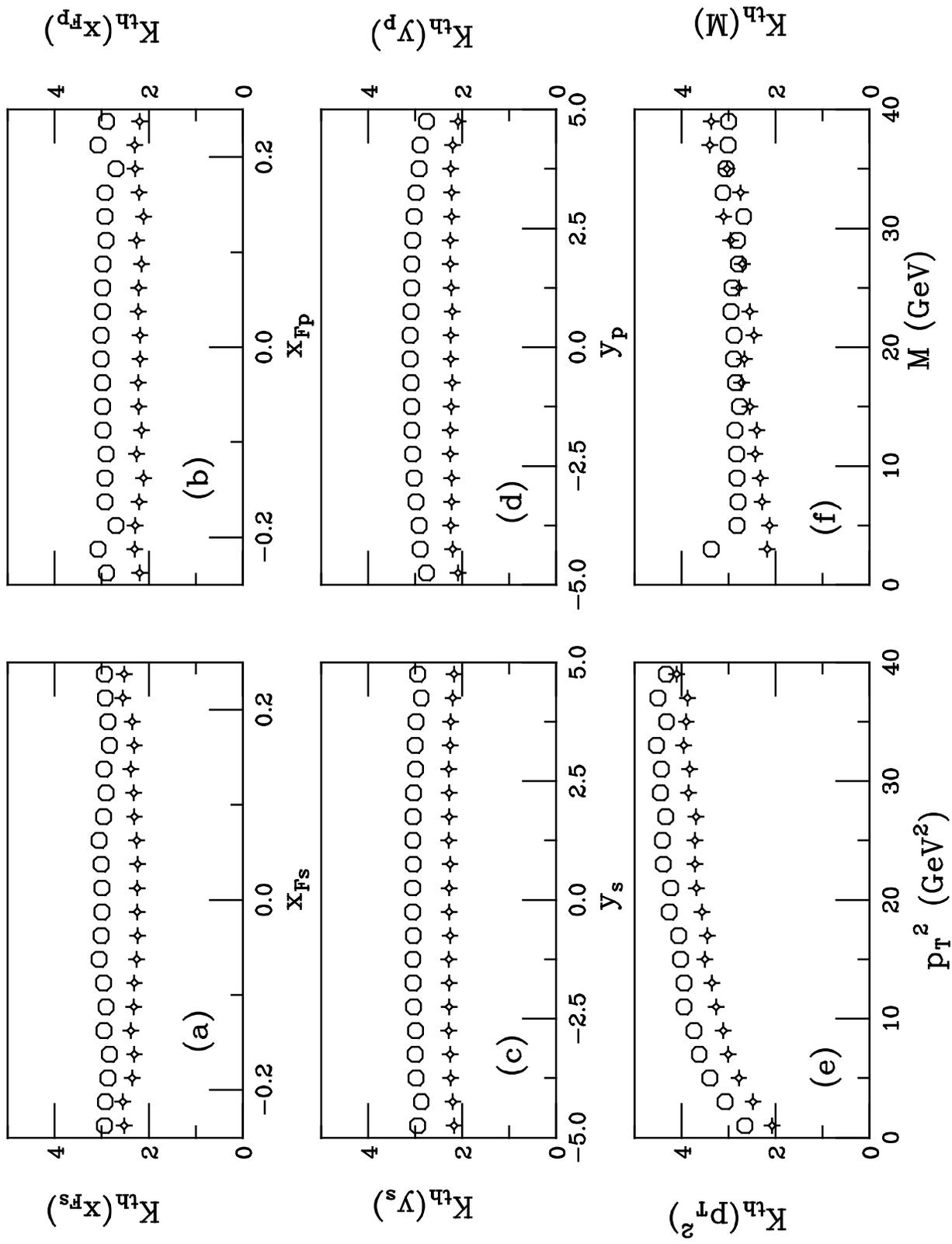
\noindent Figure 8.  The same as Fig.\ 5 for charm production at
$\sqrt{S} = 5.5$ TeV.

\vspace{0.14in}
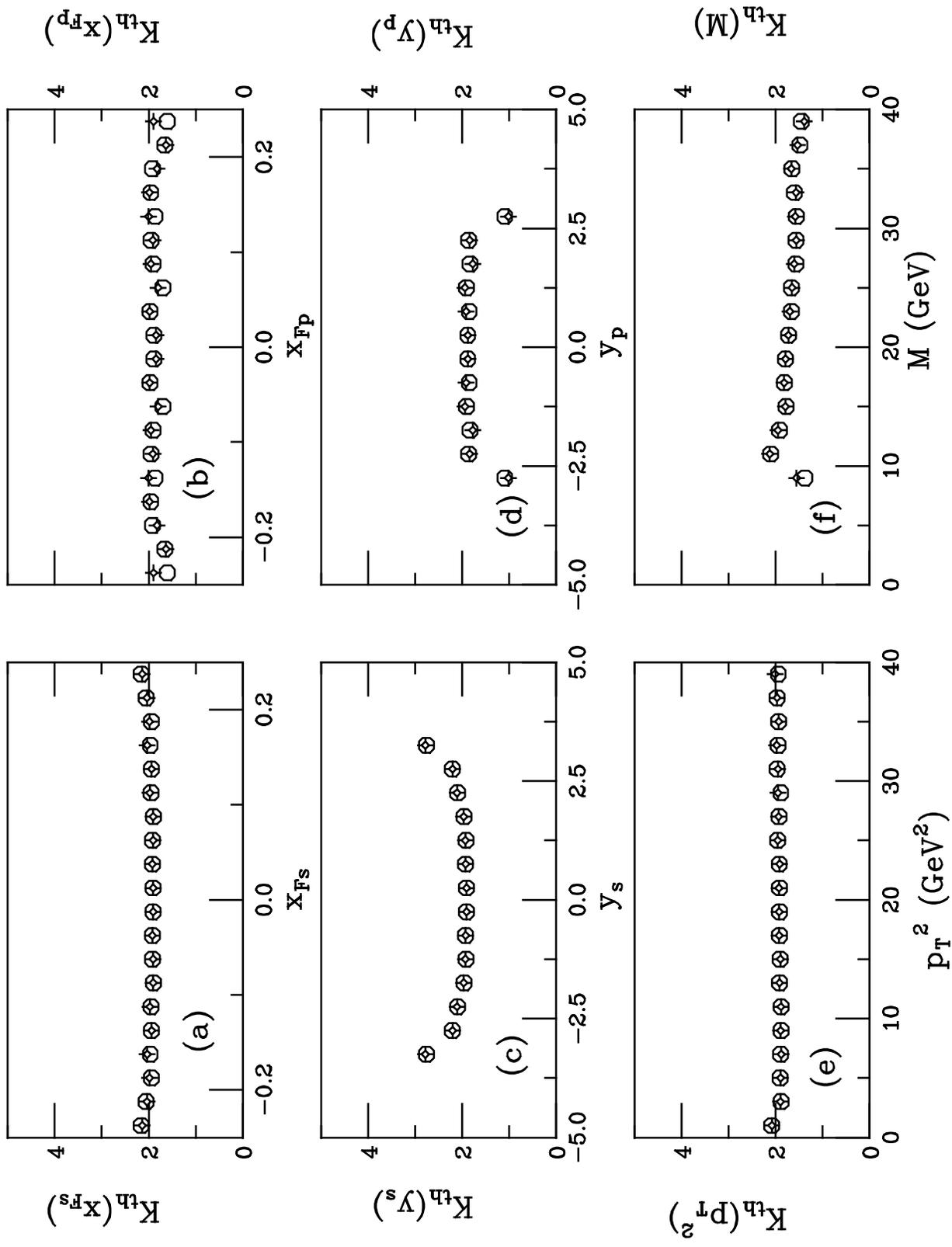
\noindent Figure 9.  The same as Fig.\ 5 for bottom production at
$\sqrt{S} = 200$ GeV.

\vspace{0.14in}
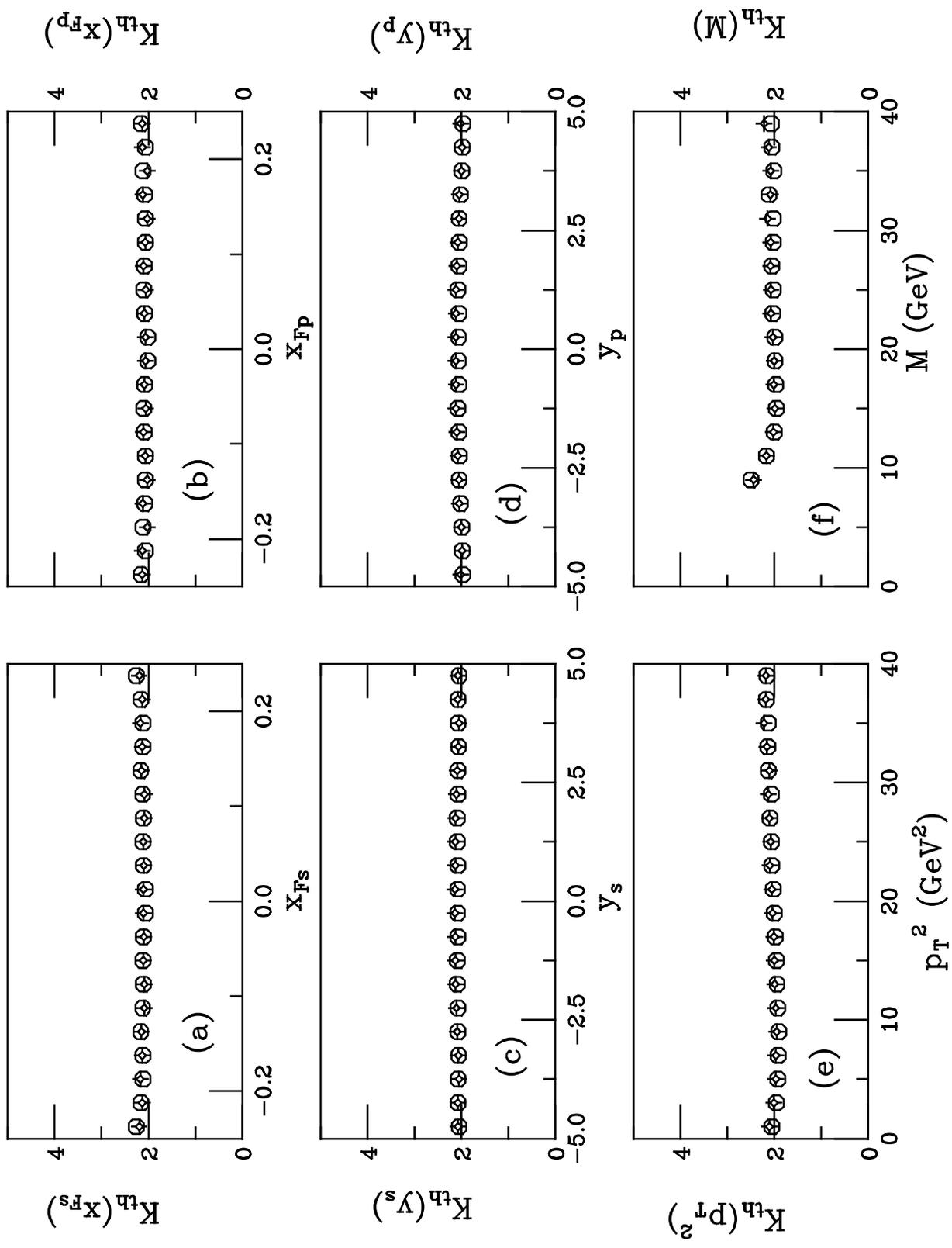
\noindent Figure 10.  The same as Fig.\ 5 for bottom production at
$\sqrt{S} = 5.5$ TeV.

\end{document}